\documentclass[vecphys]{svmult}

\usepackage{makeidx}         
\usepackage{graphicx}        
\usepackage{multicol}        

\makeindex             

\begin{document}

\title{Transporting cold atoms in optical lattices with ratchets: Symmetries and Mechanisms}
\titlerunning{Ratchets with cold atoms}
\author{S. Denisov$^{1}$, S. Flach$^{2}$, and P. H\"{a}nggi$^{1}$}
\institute{ Institut f\"ur Physik, Universit\"at  Augsburg,
       Universit\"atsstr.1, D-86135 Augsburg, Germany
\texttt{sergey.denisov@physik.uni-augsburg.de,
hanngi@physik.uni-augsburg.de} \and  Max-Planck-Institut f\"ur
Physik Komplexer Systeme, N\"othnitzer Str. 38, 01187 Dresden,
Germany \texttt{flach@mpipks-dresden.mpg.de}}

\maketitle

\section{Introduction}

Thermal fluctuations alone cannot create a steady directed transport
in an unbiased system. However, if a system is out of equilibrium,
the Second Law of Thermodynamics no longer applies, and then there
are no thermodynamical constraints on the appearance of a steady
transport \cite{Smoluchowski,Fein}. A directed current can be
generated out of a fluctuating (time-dependent) external field with
zero mean. The corresponding \textit{ratchet effect} \cite{magna,
first, adja, Reim,Astumian,Hanggi_last} has been proposed as a
physical mechanism of a microbiological motility more then a decade
ago \cite{first,adja}. Later on the ratchet idea has found diverse
applications in different areas \cite{Reim, Astumian, Hanggi_last},
from a mechanical engine \cite{Slava} up to  quantum systems  and
quantum devices \cite{Qu1, Qua2, Qu3, Qu4,Qu5,Qu6}.

When the deviation from an equilibrium regime is small (the case of
weak external fields) one may use the linear response theory in
order to estimate the answer of the system \cite{kubo,HanggiPR}.
However, due to the linearization of the response, the current value
will be strictly zero since the driving field has zero bias.
Therefore, one has to take into account nonlinear corrections and
then derive the corresponding nonlinear response functional
\cite{HanggiPR,kampen}, which may become a very complicated task, if
the nonadiabatic regime is to be considered.

To obtain a dc-current, one has to break certain discrete
symmetries, which involve simultaneous transformations in space and
time. A recently elaborated \textit{symmetry approach} \cite{Flach1,
Den} established a clear relationship between the appearance of a
directed current and broken space-time symmetries of the
\textit{equations of motion}. Thus, the symmetry analysis provides
an information about the conditions for a directed transport
appearance  without the necessity of considering a nonlinear
response functional.

Most theoretical and experimental studies have focused on  ratchet
realizations at a noisy overdamped limit \cite{Reim,Astumian,
Hanggi_last}. However, systematic studies of the underlying broken
symmetries, and the largest possible values of directed currents
achieved for different dissipation strength, show that the dc
current values typically become orders of magnitude larger in the
limit of weak dissipation \cite{Jung,disip}. The corresponding
dynamics is characterized by long space-time correlations which may
drastically increase the rectification efficiency \cite{disip, D&F}.

Fast progress in experimental studies of cold atoms ensemble
dynamics have provided clean and versatile experimental evidence of
a ratchet mechanism in the regime of weak or even vanishing
dissipation \cite{cold_rev1, cold_rev2}. The results of the
corresponding symmetry analysis for the regime of classical dynamics
has already been successfully tested with cold Rubidium and Cesium
atoms in optical lattices with a tunable weak dissipation
\cite{ren1, ren2}. Further decreasing of the dissipation strength
leads to the quantum regime \cite{cold_rev1}. Recent experiments
have shown the possibility to achieve an optical lattice with
tunable asymmetry in the quantum regime \cite{weitz}. A
Bose-Einstein condensate (BEC) loaded into an optical potential is
another candidate for a realization of  quantum ratchets in the
presence of atom-atom interactions \cite{cold_rev2}. While there is
obvious interest in experimental realizations of theoretically
predicted symmetry broken states, another important aspect of the
interface between cold atoms and the ratchet mechanism is, that new
possibilities for a control of the dynamics of atomic systems by
laser fields may be explored \cite{rice, d&k&u}.

The objective of this work is to provide a general introduction into
the symmetry analysis of the rachet effect using a simple,
non-interacting one-particle dynamics. Despite its simplicity, this
model contains all the basic aspects of classical and quantum
ratchet dynamics, and may be used also as a starting point of
incorporating atom-atom interactions.

\section{Single particle dynamics}
\label{sec:2}

We start with the simple model of an underdamped particle with mass
$m$, moving in a space-periodic potential $U(x)=U(x+\lambda)$ under
the influence of the external force $\chi(t)$ with zero mean:
\begin{equation}
m\ddot{x}+\gamma \dot{x}- f(x)-\chi(t)=0 \label{eq:particle}.
\end{equation}
Here $f(x)=-U'(x)$, $\int_{0}^{\lambda}f(x)dx=0$, and $\gamma$ is
the friction coefficient. Next, we ask whether a directed transport
with nonzero mean velocity, $\langle \dot{x} \rangle \neq0$, may
appear in the system (\ref{eq:particle}).

If $\chi(t) \equiv \xi(t)$ is a realization of a Gaussian white
(i.e. delta-correlated) noise, obeying via its correlation
properties the (second) fluctuation-dissipation theorem
\cite{HanggiPR},  Eq.(\ref{eq:particle}) then describes the thermal
equilibrium state of a particle interacting with a heat bath. From
the Second Law of Thermodynamics it is follows that a directed
transport is absent, independently of the particular choice of the
periodic potential $U(x)$ \cite{first,Reim,Jung}.

The presence of temporary correlations in $\chi(t)$ may change the
situation drastically. A simple way to get such correlations is to
use an additive periodic field $E(t)$,
\begin{equation}
\chi(t)=\xi(t)+E(t)\;,\; E(t)=E(t+T)\;,\; \int_{0}^{T}E(t)dt=0
\label{eq:noise}.
\end{equation}
If $\xi(t)$ is a realization of a white noise, then the functions
$-\xi(t)$, $\xi(t)$, and $\xi(t+\tau)$ are also realizations of the
same white noise, and their statistical weights are equal to the
statistical weight of the original realization. For what comes, the
noise term $\xi(t)$ will thus not be relevant for the following
symmetry analysis. We consider the symmetries of the deterministic
differential equation
\begin{equation}
m\ddot{x}+\gamma \dot{x}- f(x)-E(t)=0   \label{eq:part_deter}.
\end{equation}
Eq.(\ref{eq:part_deter}) contains two periodic functions, $f(x)$ and
$E(t)$, both with zero mean. The properties of the symmetries of the
Eq.(\ref{eq:part_deter}) are strongly depending on the symmetry
properties of these functions.

\section{Symmetries}
\label{sec:3}

\subsection{Symmetries of a periodic function with zero mean}
\label{sec:4}

Let us consider a periodic function $g(z+2\pi)=g(z)$ with zero mean,
$\int_{0}^{2\pi} g(z)dz=0$. This function can be expanded into a
Fourier series
\begin{equation}
g(z)=\sum_{k=-\infty}^{\infty}g_{k}\cdot \exp(ikz)
\label{eq:Fourier},
\end{equation}
where $g_{0}\equiv 0$. We will consider real-valued functions;
therefore, $g_{k}=g^{\ast}_{-k}$.

The function $g(z)$ may possess three different symmetries. First,
it can be \textit{symmetric}, $g(z+z_{0})=g(-z+z_{0})$, around a
certain argument value $z_{0}$. For such functions we will  use the
notation $g_{s}$. The Fourier expansion (\ref{eq:Fourier}) contains,
after the shift by $z_{0}$,  only cosine terms,  so
$g_{k}(z_{0})=g_{k}\cdot exp(i k z_{0})$ are  real numbers, i.e.
$g_{k}(z_{0})=g_{-k}(z_{0})$.

Second, the function $g(z)$ can be \textit{antisymmetric},
$g(z+z_{1})=-g(-z+z_{1})$, around a certain value of the argument,
$z_{1}$. For such functions we will use the notation $g_{a}$. The
corresponding Fourier expansion (\ref{eq:Fourier}) contains only
sine terms (after the shift by $z_{1}$), so $g_{k}(z_{0})=g_{k}\cdot
exp (i k z_{0})$ are pure imaginary numbers, and
$g_{k}(z_{0})=-g_{-k}(z_{0})$.

Finally, the function $g(z)$ can be \textit{shift-symmetric},
$g(z)=-g(z+\pi)$. The Fourier expansion of a shift-symmetric
function $g_{sh}(z)$ contains odd harmonics only, $g_{2m} \equiv 0$.

It is straightforward to show that a periodic function $g(z)$ can
have either none of the above mentioned symmetries, or exactly one
of them, or all three of them. Let us consider several simple
examples. The function $cos(z)$ possesses all three symmetries. The
function $cos(z)+cos(3z+\phi)$ always possesses shift-symmetry and
in addition may be simultaneously symmetric and antisymmetric for
$\phi=0, \pm \pi$. The function $cos(z)+cos(2z+\phi)$ is not
shift-symmetric, thus it will either have no other symmetry at all,
except for $\phi=0, \pm \pi$ (symmetric), and $\phi=\pm \pi/2$
(antisymmetric).

\subsection{Symmetries of the equations of motion}
The system  dynamics in eq. (\ref{eq:part_deter}) can be described
by three first-order autonomous differential equations,
\begin{equation}
\dot{x}=\frac{p}{m}\;,\;\dot{p}= f(x)+E(\tau)-\frac{\gamma}{m}
p\;,\; \dot{\tau}=1\;. \label{eq:three_eqs}
\end{equation}
The phase-space dimension is three. We are looking for symmetry
transformations, $\widehat{S}$, which do not change the equation
(\ref{eq:three_eqs}), but do change the sign of the velocity
$\dot{x}$. Such transformations map the phase space $\{x,p,\tau\}$
onto itself. If we find such a transformation, we then apply it to
all points of a given trajectory. We get a new manifold in phase
space, which  also represents a trajectory, i. e. a solution of the
equations (\ref{eq:three_eqs}). The original trajectory and its
image may coincide (or may not).

Let us assume that we have found such a transformation. Next, we
consider the mean velocity, $\bar{v}=\lim_{s \rightarrow \infty}
\frac{1}{s}(x(t_{0}+s)-x(t_{0}))$, on the original trajectory. If
the trajectory and its image coincide, then $\bar{v}=0$. If they are
different then their velocities have the same absolute value but
opposite signs. If, in addition, both the trajectories have the same
statistical weights in the presence of a white noise, then we can
conclude that the average current in the system
(\ref{eq:part_deter}) is equal to zero \cite{Flach1}.

There are only two possible types of transformations which change
the sign of the velocity $\dot{x}$: they include either a
time-reversal operation, $t \rightarrow -t$, or a space inversion,
$x \rightarrow -x$ (but not both operations simultaneously!).

The following symmetries can be identified \cite{Flach1}:
\begin{eqnarray}
& \, & \hat{S}_a\;\;: \;\; x \rightarrow -x\;\;, \;\; t\rightarrow
t+\frac{T}{2}\;\;, \;\; {\rm if} \; \{f_{a}\;,\; E_{sh} \} \;\; ;
\nonumber
\\
& \, & \hat{S}_b \;\;: \;\; x \rightarrow x\;\;, \;\; t\rightarrow
-t \;\; , \;\; {\rm \qquad \ if} \; \{E_s\;,\;\gamma = 0\} \;\; ;
\label{sym1}
\\
& \, & \hat{S}_c \;\; : \;\; x \rightarrow x+\frac{\lambda}{2} \;\;,
\;\; t \rightarrow -t \;\; , \;\; {\rm \ if} \;
\{f_{sh}\;,\;E_a\;,\;m=0\} \;\; . \nonumber
\end{eqnarray}
The symmetry $\hat{S}_b$ requires zero dissipation, $\gamma=0$, i.e.
it requires the Hamiltonian regime, and the symmetry $\hat{S}_c$ can
be fulfilled in the overdamped limit (i.e. $m=0$) only. Note that
all symmetries require certain symmetry properties of the function
$E(t)$. Usually, an experimental setup allows to tune the shape of
the time-dependent field $E(t)$ easier than the shape of the
spatially periodic potential \cite{Marchesoni}. A proper choice of
the force $E(t)$ may break all three symmetries for any coordinate
dependence of the force $f(x)$. We restrict the further
consideration to the case of a symmetric potential $U(x)=1-\cos(x)$
while using a bi-harmonic driving force,
\begin{equation}
E(t)=E_{1}cos(t)+E_{2}cos(2t+\theta), \label{biharm}
\end{equation}
for a symmetry violation. If $\theta \neq 0, \pi/2, \pi, 3\pi/2$
then all three symmetries (\ref{sym1}) are broken and we may count
on a nonzero mean velocity, $v \neq 0$.

\subsection{The case of quasiperiodic functions}

We  generalize the symmetry analysis to the case of quasiperiodic
driving field $E(t)$ \cite{Pikovsky, quasi}.

We consider a quasiperiodic function $g(z)$ to be of the form
\begin{equation}
g(z)\equiv \tilde{g}(z_1,z_2,...,z_N)\;,\;\frac{\partial
z_{i}}{\partial z}= \Omega_i \label{3-1}
\end{equation}
where all ratios $\Omega_i/\Omega_j$ are irrational if $i \neq j$
and $\tilde{g}(z_1,z_2,...,z_i+2\pi,...,z_N) =
\tilde{g}(z_1,z_2,...,z_i,...,z_N)$ for any $i$. Such a function may
have numerous symmetries. With respect to the following symmetry
analysis of the equation of motion we will list here only those
symmetries of $\tilde{g}$ which are of relevance. It can be
symmetric $\tilde{g}_s(z_1,z_2,...,z_N)=
\tilde{g}_s(-z_1,-z_2,...,-z_N)$, antisymmetric
$\tilde{g}_a(z_1,z_2,...,z_N)= -\tilde{g}_a(-z_1,-z_2,...,-z_N)$. It
can be also shift-symmetric for a given set of indices
$\tilde{g}_{sh,\{i,j,...,m\}}$ which means that $\tilde{g}$ changes
sign when a shift by $\pi$ is performed in the direction of each
$z_i,z_j,...,z_m$ only, leaving the other variables unchanged.

The relevant symmetry properties of $\tilde{g}$ are thus studied on
the compact space of variables $\{ z_1,z_2,...,z_N \}$. The
irrationality of the frequency ratios guarantees that in the course
of evolution of $z$ this compact space is densely  scanned by these
variables with uniform density in the limit of large $z$. At the
same time we note that it is always possible to find a large enough
value $Z$ such that
\begin{equation}
\lim_{\tau \rightarrow \infty} \frac{1}{\tau} \int_0^{\tau}
(g(z+Z)-g(z))^2 {\rm d}z \; < \; \epsilon \label{3-1-1}
\end{equation}
with (arbitrarily) small absolute value of $\epsilon$. For a given
value of $\epsilon$ this defines a quasiperiod $Z$ of the function
$g(z)$.

In order to make the symmetry analysis of the equation of motion
transparent, we rewrite it (skipping the noise term) in the
following form \cite{quasi}:
\begin{eqnarray}\label{3-2}
m\ddot{x} + \gamma \dot{x} - f(x) - E(\phi_1,\phi_2,...,\phi_N) =0
\;,\\
\dot{\phi}_1 = \omega_1 \;\;\;,\nonumber \\
\dot{\phi}_2 = \omega_2 \;\;\;, \nonumber \\
. \nonumber \\
.\nonumber \\
.\nonumber \\
\dot{\phi}_N = \omega_N \;\;. \nonumber
\end{eqnarray}
The function $f(x)$ is also assumed to be quasiperiodic with $M$
corresponding spatial frequencies.

The following symmetries can be identified, which change the sign of
$\langle \dot{x} \rangle$ and leave (\ref{3-2}) unchanged:
\begin{eqnarray}
& \, \tilde{S}_a\;\;: x \rightarrow -x\;, \;
\phi_{i,j,...,m}\rightarrow \phi_{i,j,...,m}+\pi\;, \; {\rm if} \;
\{f_{a}\;,\; E_{sh,\{i,j,...,m\}} \}\;,
~~\nonumber
\\
\, & \tilde{S}_b \;\;: \;\; x \rightarrow x\;\;, \;\; t\rightarrow
-t \;\; , \;\; {\rm \qquad \ if} \; \{E_s\;,\;\gamma = 0\} \;\;
;~~~~~~~~~~~~~~~~~~~~~~ \label{symqp1}
\\
\, & \tilde{S}_c \;\; : \;\; x \rightarrow x+\frac{\lambda}{2} \;\;,
\;\; t \rightarrow -t \;\; , \;\; {\rm \ if} \;
\{f_{sh,\{1,2,3,...,M\}}\;,\;E_a\;,\;m=0\} \;\; . \nonumber
\end{eqnarray}
The symmetry $\tilde{S}_a$ is actually a set of various symmetry
operations which are defined by the given subset of indices
$\{i,j,...,m\}$.

The prediction then is, that if for a given set of parameters any of
the relevant symmetries (\ref{symqp1}) is fulfilled, the average
current will be zero. If however the choice of functions $f(x)$ and
$E(t)$ is such that the symmetries are violated, a nonzero current
is expected to emerge.

\section{Dynamical mechanisms of rectification: The Hamiltonian limit}
\label{sec:5}

Let us consider the limit $\gamma=0$ (Hamiltonian case)
\cite{Flach1, D&F}. Due to time and space periodicity of the system
(\ref{eq:part_deter}) we can map the original three-dimensional
phase space $(x, p, t)$ onto a two-dimensional cylinder,
$\texttt{T}^{2}=(x \; mod  1,p)$, by using the stroboscopic
Poincar$\acute{e}$ section after each period $T=2\pi/\omega$. For
given initial conditions $\{x(0),p(0)\}$, we integrate the system
over time $T$, and then plot the final point, $\{x(T),p(T)\}$, on
the cylinder $\texttt{T}^{2}$.

\begin{figure}[t]
\centering
\includegraphics*[width=.6\textwidth, angle=-90]{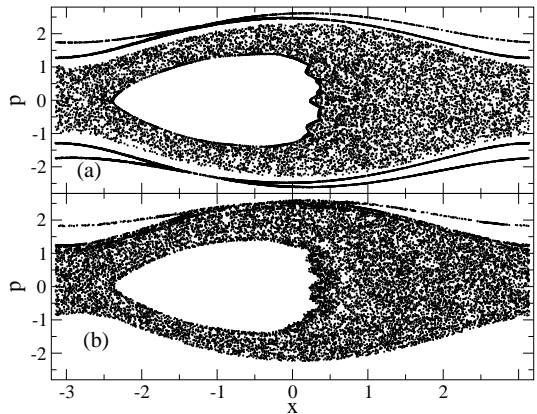}
\caption[]{Poincar\'{e} map for the system
(\ref{eq:three_eqs},\ref{biharm}). The parameters are $E_{1}=0.252$,
$E_{2}=0.052$, $\gamma=0$. (a) $\theta=0$; (b) $\theta=\pi/2$. }
\label{fig:1}
\end{figure}

\begin{figure}[t]
\centering
\includegraphics*[width=.6\textwidth, angle=-90]{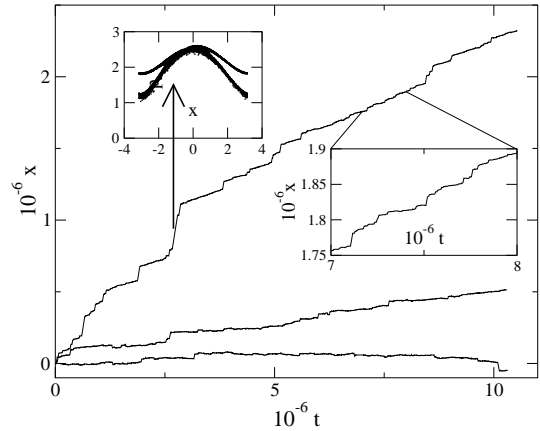}
\caption[]{ $x(t)$ for $\theta = 0, \pi/5, \pi/2$ (lower, middle and
upper curves, respectively). Left upper inset: Poincar\'{e} map for
a single ballistic flight, $\theta=\pi/2$. Right inset: zoom of
$x(t)$ for the case $\theta=\pi/2$.} \label{fig:2}
\end{figure}

For $E(t)=0$ the system (\ref{eq:part_deter})  is integrable and
there is a separatrix in the phase space which separates oscillating
and running solutions. A non-zero field $E(t)$ destroys the
separatrix and leads to the appearance of a stochastic layer (see
Fig.\,\ref{fig:1}). In this part of the phase space the system
dynamics is ergodic, i. e. all average characteristics are the same
for all trajectories, launched inside the layer. Therefore, the
symmetry analysis is valid for all trajectories on this manifold.
Numerical studies have confirmed this conclusion \cite{Flach1,D&F,
Den}. Fig.\,\ref{fig:2} shows several trajectories $x(t)$ from
chaotic layers and illustrates the fact that the violation of
symmetries causes a directed motion of the particle.

The dynamics within the stochastic layer can be roughly subdivided
into two distinct fractions. The first one corresponds to ballistic
flights near the layer boundaries. They appear due to a sticking
effect \cite{Ham_cl}. A random diffusion within a chaotic  bulk is
attributed to the second fraction. A rectification effect appears
due to a violation of the balance between ballistic flights in
opposite directions \cite{D&F}. This interpretation supports the
view, that even in the presence of damping and noise, the ratchet
mechanism relies on harvesting on temporal correlations of the
underlying dynamical system. Ballistic flights are just such
examples of long temporal correlations on a trajectory which is
overall chaotic. Therefore it is not surprising, that the ratchet
effect is stronger in the dissipationless limit, since dissipation
will introduce finite (and possibly short) time scales which cut the
temporal correlations down. The averaged drift velocity can be
estimated by using a sum rule \cite{Ketz}. From the corresponding
approach, which is based on a statistical argument by the authors of
Ref. \cite{Ketz}, it  follows, that a mixed space, i.e. a stochastic
layer with boundaries and embedded regular submanifolds (islands),
presents the necessary condition for a directed transport.

The adding of a non-zero dissipation, $\gamma \neq 0$, does not
change the situation drastically \cite{Den}. The symmetry analysis
is still valid for this case. The phase space is shared by different
transporting and non-trasnporting attractors with their
corresponding basins of attraction, which are strongly entangled
inside the former stochastic layer region. A symmetry violation
causes a desymmetrization of basins.  Finally, a weak noise leads to
a trajectory wandering over different basins, sticking to
corresponding attractors, and, finally, to the rectification effect.
The long flights which appear at the Hamiltonian limit are damped
after a characteristic time which is the shorter, the larger the
dissipation strength $\gamma$ is \cite{D&F, Den}.

A systematic analysis shows that, under the condition of full
symmetry violation, the approach of the dissipationless limit leads
to a drastic increase of the dc current value \cite{disip}, which
depends on the characteristics of the stochastic layer \cite{D&F}.
It has been shown that, in a full accordance with the symmetry
analysis, the dc current disappears near $\theta=0,\pi$ for the case
of weak dissipation, and near $\theta=\pm/2$ at the strong
dissipation limit. The value of the phase $\theta$, at which the
current becomes zero, is a monotonous function of the dissipation
strength $\gamma$ \cite{disip}.

An inclusion of a  dc-component to  the external field,
$\tilde{E}(t)=E(t)+E_{dc}$, may lead to a directed transport against
a constant bias $E_{dc}$, even in the Hamiltonian limit \cite{bias}.

The abovementioned results have been confirmed in cold atoms
experiments, performed in the group of Renzoni \cite{ren1}. In these
experiments, atoms of Cs and Rb have been cooled  to temperatures of
several mK. An optical standing wave, created by a pair of
counter-propagating laser beams, formed a periodic potential for the
atoms. Finally, a time-dependent force $E(t)$ has been introduced
through a periodic modulation of the phase for one of the beams. The
results of the above symmetry analysis have been verified by
changing the  relative phase $\phi$  and by tuning the effective
dissipation strength.

The case of the quasiperiodic driving force $E(t)$ for cold atoms
ratchets also has been studied experimentally \cite{ren2}, with a
similar outcome.

\section{Resonant enhancement of transport with quantum ratchets}
\label{sec:6} A quantum extension of the (dissipationless) system
dynamics in eq. (\ref{eq:part_deter}) can readily be achieved
\cite{our_quant}.  The system evolution can be described by the
Schr\"{o}dinger equation,
\begin{equation}
i \hbar \frac{\partial}{\partial t} |\psi(t)\rangle =
H(t)|\psi(t)\rangle, \label{eq:Schrodin}
\end{equation}
where the  Hamiltonian $H$ is of the form
\begin{equation}
H(x,p,t)=\frac{p^{2}}{2}+(1+\cos(x))- x E(t). \label{eq:ham}
\end{equation}
The system (\ref{eq:Schrodin}) describes a cloud of noninteracting
atoms, placed into a periodic potential (formed by two
counter-propagating laser beams) and exposed to an external ac field
(\ref{biharm})\footnote{The dissipation may be included into quantum
dynamics by coupling the system (\ref{eq:ham}) to a heat bath,
$H_{diss}(x,p,t,\{\mathbf{q}\})=H(x,p,t)+H_{B}(x,\{\mathbf{q}\})$.
Here $H_{B}(x,\{\mathbf{q}\})$ describes  an ensemble of harmonic
oscillators $\{\mathbf{q}\}$ at thermal equilibrium  interacting
with the system \cite{Qu1}.}.

Because of the time and space periodicity of the Hamiltonian
(\ref{eq:ham}), the solutions  $|\psi_{\alpha}(t+t_0)\rangle =
U(t,t_{0})|\psi_{\alpha}(t_0)\rangle$ of the Schr\"{o}dinger
equation (\ref{eq:Schrodin}) can be characterized by the
eigenfunctions of the Floquet operator $U(T,t_0)$ which satisfy the
Floquet theorem: $ |\psi_{\alpha}(t)\rangle=
\exp(-i\frac{E_{\alpha}}{T}t) |\phi_{\alpha}(t)\rangle$,
$|\phi_{\alpha}(t+T)\rangle=|\phi_{\alpha}(t)\rangle$ (here $t_{0}$
is the initial time). The quasienergies $E_{\alpha}$ $(-\pi <
E_{\alpha} < \pi)$ and the Floquet eigenstates can be obtained as
solutions of the eigenvalue problem of the Floquet operator
\begin{equation}
U(T,t_{0})|\phi_{\alpha}(t_{0})\rangle = e^{-i
E_{\alpha}}|\phi_{\alpha}(t_{0})\rangle \label{eq:Floquet}
\end{equation}
with $\alpha$ denoting the band index and with $k$ being the wave
vector \cite{our quant, Grif,Gong}. An initial state can be expanded
over Floquet-Bloch eigenstates, $|\psi(t_{0})=$ $\sum_{\alpha, k}
C_{\alpha,k}(t_{0}) |\phi_{\alpha,k}\rangle$ and the  subsequent
state's evolution is encoded in the coefficients $\{C_{\alpha,
k}\}$.  We restrict further consideration to the case $\kappa=0$
which corresponds to initial states where atoms equally populate all
(or many) wells of the spatial potential.

\begin{figure}[t]
\begin{tabular}{cc}
\includegraphics[width=3.5cm,height=4cm]{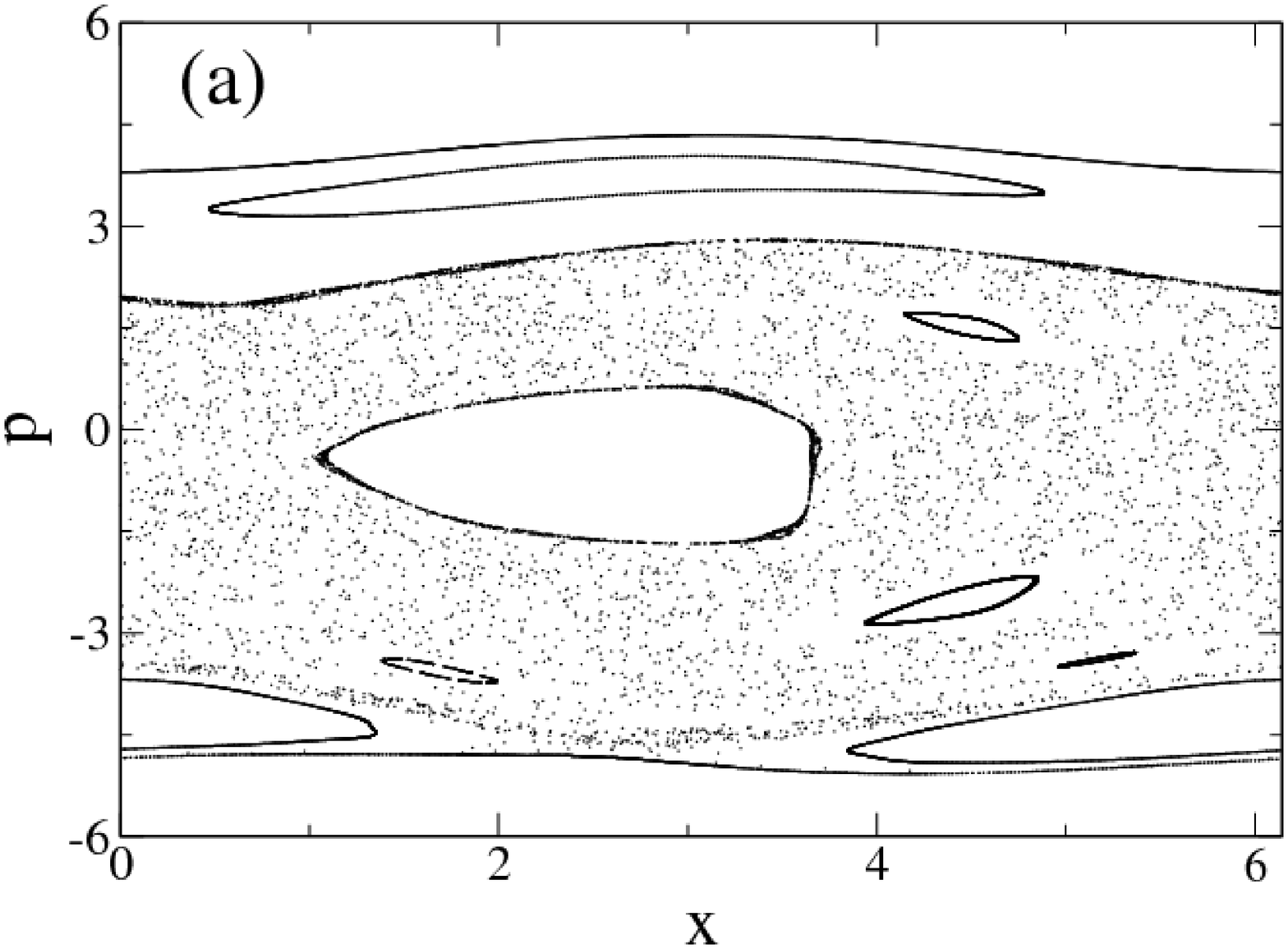}
\includegraphics[width=4.2cm,height=4cm]{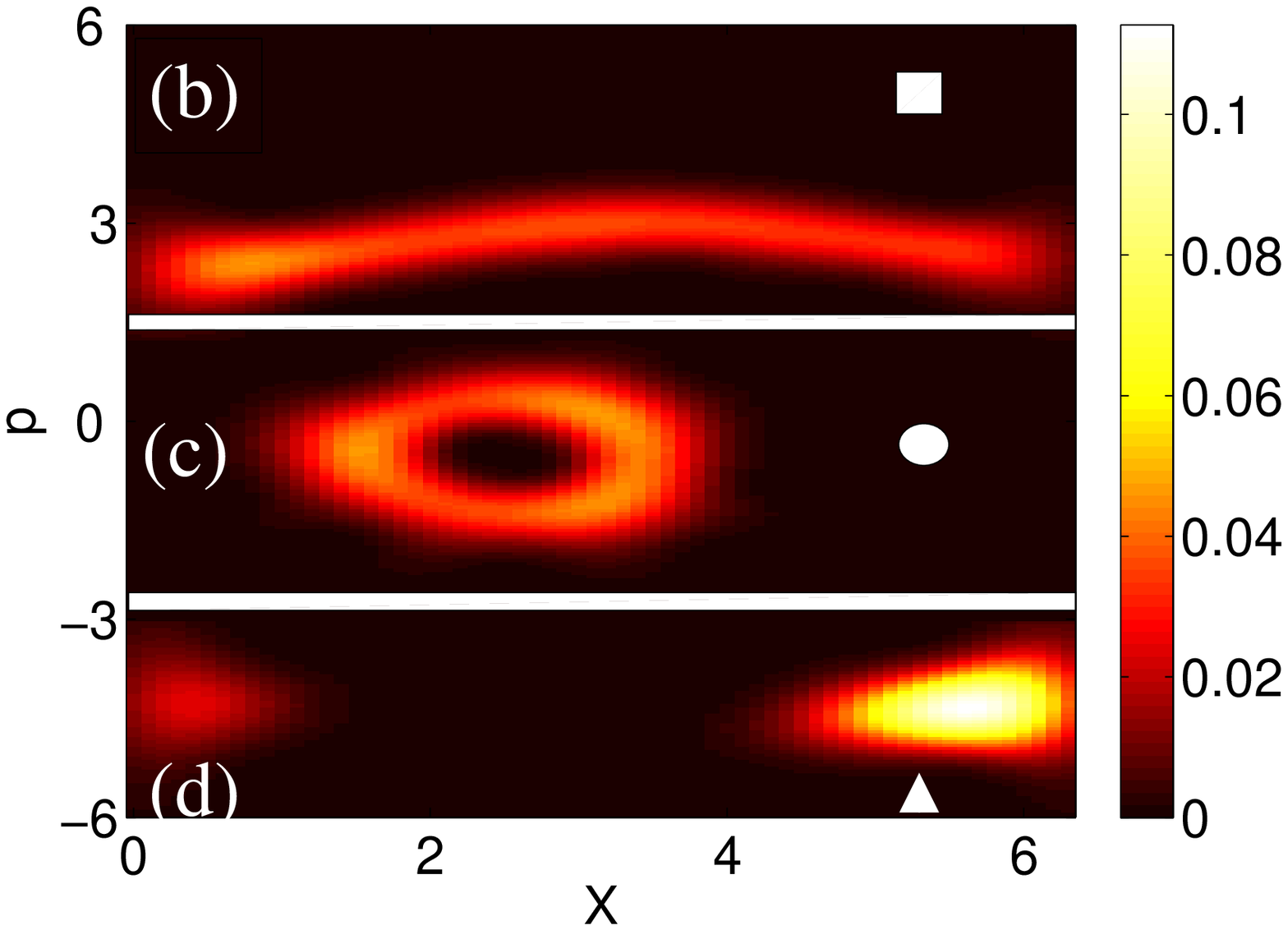}
\end{tabular}
\begin{tabular}{cc}
\includegraphics[width=4cm,height=4cm]{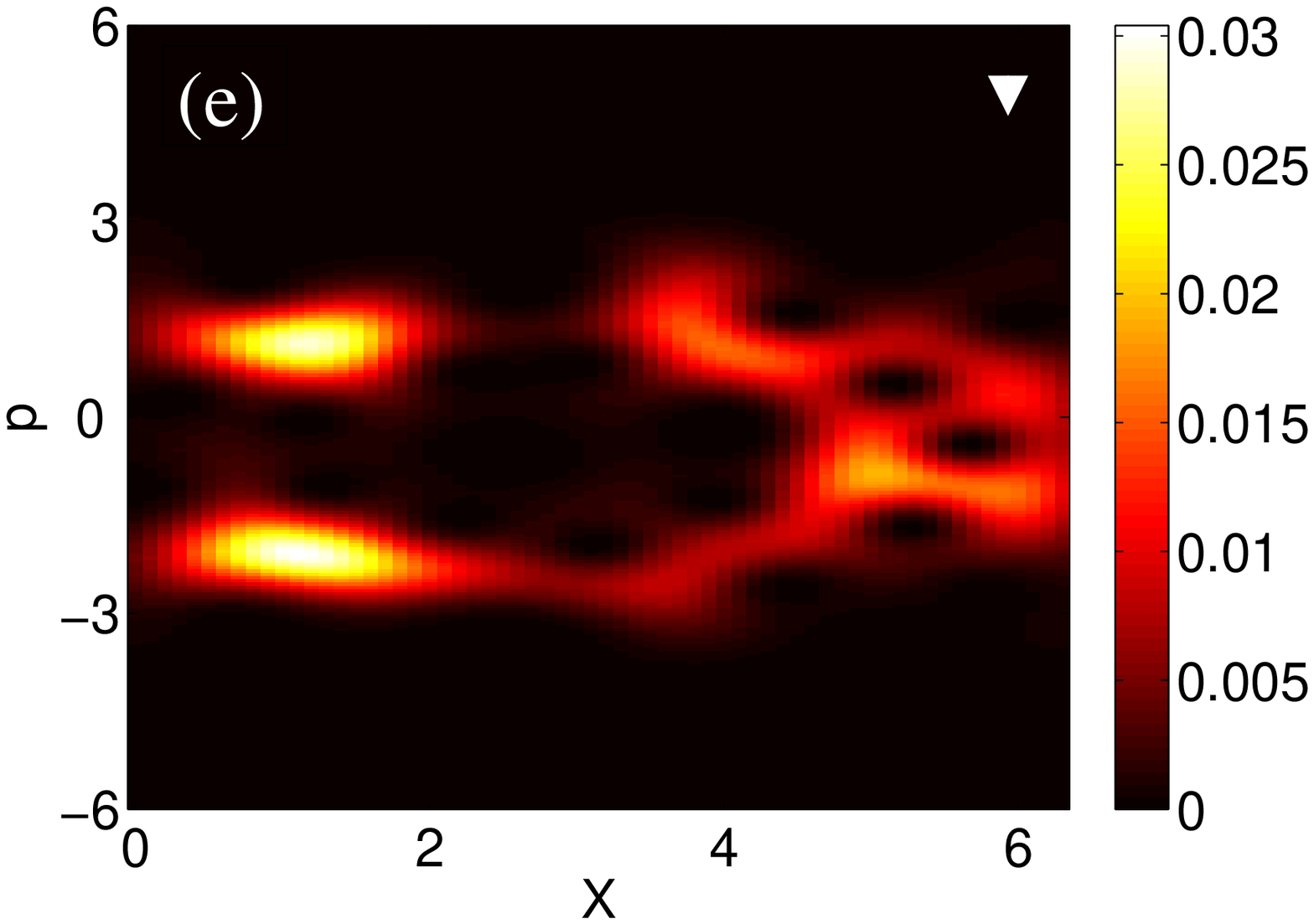}
\includegraphics[width=4cm,height=4cm]{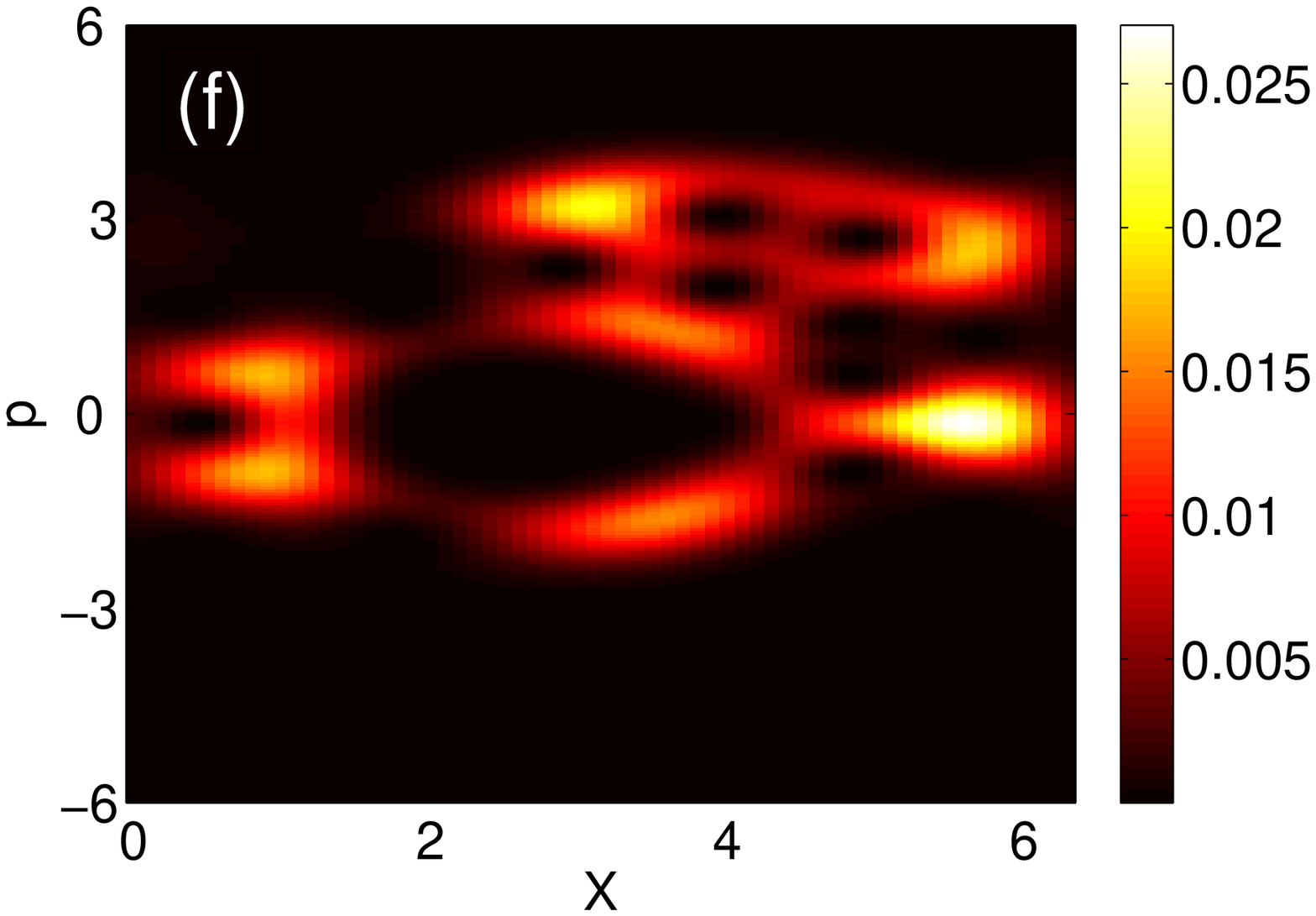}
\end{tabular}
\label{Figure1} \caption{(Color online) (a) Poincar\'{e} section for
the classical limit, (\ref{biharm}, \ref{eq:ham}); (b-f) Husimi
representations for different Floquet eigenstates for the
Hamiltonian (\ref{eq:ham}) with $\hbar=0.2$ (momentum is in units of
the recoil momentum, $p_{r}=\hbar k_{L}$, with $k_{L}=1$). The
parameters are $E_{1}=E_{2}=2$, $\omega=2$, $\theta=-\pi/2$ and
$t_{0}=0$ for (b-e), and $E_{1}=3.26$, $E_{2}=1$, $\omega=3$,
$\theta=-\pi/2$ and $t_{0}=0$ for (f). } \label{Fig:husi}
\end{figure}

The mean momentum expectation value,
\begin{equation}
J(t_0)= \lim_{t\rightarrow \infty} 1/t \int_{t_0}^{t}\langle
\psi(t,t_0)|\hat{p}|\psi(t,t_0)\rangle, \label{quant_curr}
\end{equation}
measures the asymptotic current. Expanding the wave function over
the Floquet states the current becomes
\begin{equation}
J(t_0)=\sum_{\alpha} \langle p \rangle_{\alpha}
|C_{\alpha}(t_{0})|^{2}, \label{eq:current}
\end{equation}
where $\langle p \rangle_{\alpha}$ is the mean momentum of the
Floquet state $|\phi_{\alpha}\rangle$  \cite{our_quant,Gong}.

The analysis of the transport properties of the eigenstates shows
that the quantum system  inherits the symmetries of its classical
counterpart \cite{our_quant}. In particular, the symmetries of the
classical equations of motion translate into symmetries of the
Floquet operator. The presence of any of these symmetries results in
a vanishing the time-averaged expectation value of the momentum
operator for each Floquet eigenstate: $\langle p
\rangle_{\alpha}=0$. Thus, if one of the symmetries, $\tilde{S}_a$,
$\tilde{S}_b$ (\ref{symqp1}), holds then $\langle p
\rangle_{\alpha}=0$ for all $\alpha$. Consequently $J(t_0)=0$ in
this case.

By using the Husimi representation \cite{Husimi} we can visualize
different eigenstates in the phase space, $\{x,p,\tau\}$ and
establish a correspondence between them and the mixed phase space
structures for the classical limit (Fig.\,\ref{Fig:husi}).

\begin{figure}[t]
\includegraphics*[width=.8\textwidth, angle=0]{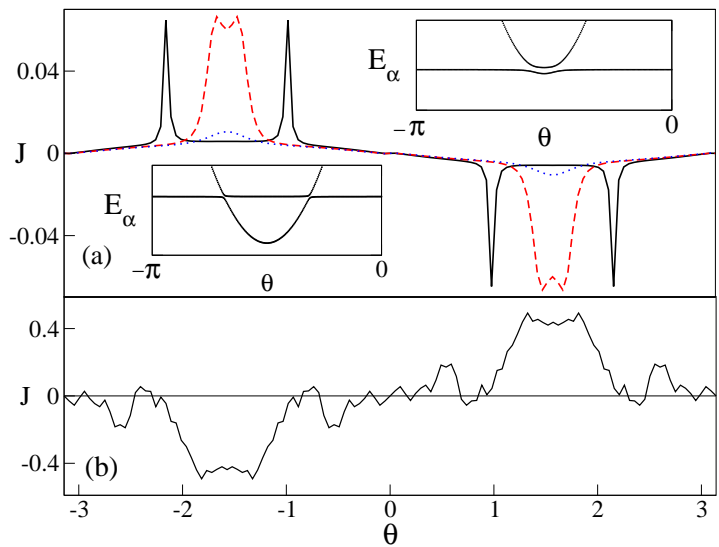}
\caption{ (a) The average current $J$ (in units of the recoil
momentum) vs $\theta$ for different amplitude values of the second
harmonic, $E_{2}$: $0.95$ (pointed line), $1$ (dashed line) and
$1.2$ (solid line). Insets: relevant details of the quasienergy
spectrum versus $\theta$ in the resonance region for $E_{2}=1$ (top
right) and $E_{2}=1.2$ (bottom left). The parameters are
$E_{1}=3.26$ and $\omega=3$. (b) The average current $J$ (in units
of the recoil momentum) vs $\theta$ for $E_{1}=3$, $E_{2}=1.5$ and
$\omega=1$.} \label{Fig:current}
\end{figure}

Since the Schr\"{o}dinger equation (\ref{eq:Schrodin}) is linear,
the system  maintains a memory of the initial condition for infinite
times \cite{Qchaos}. The asymptotic current value depends on the
initial time, $t_{0}$, and on the initial wave function,
$\psi(t_{0})$. For a given initial wave function, $|\psi\rangle
=|0\rangle $, we can assign a unique current value by performing an
averaging over the initial time $t_{0}$, $J=1/T
\int_{0}^{T}J(t_{0})dt_{0}$ \cite{our_quant}. Fig.\ref{Fig:current}
shows the dependence of the average current on the asymmetry
parameter $\theta$. Sharp resonant peaks for $E_2=1.2$ where the
current value changes drastically are associated with interactions
between two different Floquet eigenstates. The Husimi distributions
 show that one state locates in the chaotic layer, and
another one in a transporting island. Off resonance the initial
state mainly overlaps with the chaotic state, which  yields some
nonzero, yet small, current. In resonance Floquet states mix, and
thus the new eigenstates contain contributions both from the
original chaotic state as well as from the regular transporting
island state. The Husimi distribution of the mixed state is shown in
Fig.\ref{Fig:husi}(f), the strong asymmetry is clearly observed. The
regular island state has a much larger current contribution,
resulting in a strong enhancement of the current.

To conclude this section, we would like to emphasize the following
two points. For both cases, i.e. the classical and the quantum one,
the overall, total current over the whole momentum space is zero
\cite{Ketz}. Thus, it is essential to have the initial state
prepared  localized near the line $p=0$, because for broad initial
distributions the asymptotic current tends to zero. However, if the
dynamics is restricted to the lowest band of the periodic potential,
no current rectification does occur \cite{Goy}.

\section{Outlook}
\label{sec:7}

This surveyed symmetry analysis, originally put forward in Refs.
\cite{Flach1, Den}, provides a general toolbox for the  prediction
of dynamical regimes for which one can (or cannot) obtain the
rectification and directed current phenomenon for a given transport
dynamics. First, one has to set up the equations of motions and
define a the observable (current, magnetization, angular velocity,
energy flux, etc) which should become nonzero, in terms of these
dynamical variables. Then, one examines whether there exist
transformations (symmetries) which change the sign of the observable
and at the same time leave the equations of motion invariant. Upon
breaking all the symmetries one can expect the emergence of a
non-zero, directed current. This strategy has been successfully
tested with Josephson junctions (fluxon directed motion)
\cite{soliton, Ustinov} and as well with paramagnetic resonance
experiments (spin magnetization by a zero-mean field) \cite{spin,
arim}.

Herein, we focused only on the one-dimensional case. By use of  more
laser beams, experimentalists can  fabricate two- and
three-dimensional optical potentials \cite{cold_rev2}. By changing
the relative phase between  lattice beams, 2D- and 3D-potentials
with different symmetries and topologies can be achieved
\cite{2D3D}. This fact incites for an extension of the present
ratchet studies into higher dimensions.

Moreover, for the phenomenon of Bose-Einstein-condensation (BEC) of
cold gases, interactions between atoms become essential and
nonlinearities start to play an important role \cite{cold_rev2}.
Many features of BEC dynamics are manifestations of general concepts
of nonlinear physics, such as soliton creation and propagation.
These collective excitations can then themselves be subjected to a
ratchet transport mechanism \cite{soliton1}.

\printindex
\end{document}